\documentclass[utf8]{frontiersSCNS} 

\usepackage{url,hyperref,lineno,microtype,subcaption}
\usepackage[onehalfspacing]{setspace}

\def\keyFont{\fontsize{8}{11}\helveticabold }
\def\firstAuthorLast{Pouke {et~al.}} 
\def\Authors{Matti Pouke\,$^{1,*}$, Katherine J. Mimnaugh\,$^{1}$, Alexis Chambers\,$^{1}$, Timo Ojala\,$^{1}$ and Steven M. LaValle\,$^{1}$}
\def\Address{$^{1}$Center for Ubiquitous Computing, Faculty of Information Technology and Electrical Engineering, University of Oulu, Oulu, Finland} 
\def\corrAuthor{Matti Pouke}

\def\corrEmail{matti.pouke@oulu.fi}

\begin{document}
\onecolumn
\firstpage{1}

\title[Plausibility Paradox for Resized Users]{The Plausibility Paradox For Resized Users in Virtual Environments} 




\author[\firstAuthorLast ]{\Authors} 
\address{} 
\correspondance{} 

\extraAuth{}

\maketitle

\begin{abstract}

\section{}
This paper identifies and confirms a perceptual phenomenon: when users interact with simulated objects in a virtual environment where the users' scale deviates greatly from normal, there is a mismatch between the object physics they consider realistic and the object physics that would be correct at that scale. We report the findings of two studies investigating the relationship between perceived realism and a physically accurate approximation of reality in a virtual reality experience in which the user has been scaled by a factor of ten. Study 1 investigated perception of physics when scaled-down by a factor of ten, whereas Study 2 focused on enlargement by a similar amount. Studies were carried out as within-subjects experiments in which a total of 84 subjects performed simple interaction tasks with objects under two different physics simulation conditions. In the \textit{true physics} condition, the objects, when dropped and thrown, behaved accurately according to the physics that would be correct at that either reduced or enlarged scale in the real world. In the \textit{movie physics} condition, the objects behaved in a similar manner as they would if no scaling of the user had occurred. We found that a significant majority of the users considered the \textit{movie physics} condition to be the more realistic one. However, at enlarged scale, many users considered \textit{true physics} to match their expectations even if they ultimately believed \textit{movie physics} to be the realistic condition. We argue that our findings have implications for many virtual reality and telepresence applications involving operation with simulated or physical objects in abnormal and especially small scales. 


\tiny
 \keyFont{ \section{Keywords:} Virtual Reality, Perception, Scaling, Plausibility, Human Factors} 
\end{abstract}

\section{Introduction}
Many studies have confirmed the so-called "body-scaling effect": if presented with mismatching size cues, humans tend to use their visible body as the dominant cue when perceiving sizes and distances (\citealt{banakou_illusory_2013, langbehn_scale_2016, linkenauger_welcome_2013, van_der_hoort_being_2011, ogawa2017distortion}). For example, if a person was somehow shrunk to the size of a doll, the person would be inclined to regard the world as scaled-up and him/herself as normal-sized (\citealt{van_der_hoort_being_2011}). In this paper, we investigate the human perception of physics, specifically when subjects have been either scaled down or up by a significant amount. We believe this relatively underrepresented topic has implications to various virtual reality (VR) and telepresence applications. More specifically, we focus on the subjective credibility of rigid body dynamics when subjects are presented with realistic and unrealistic approximations of object motions when either scaled down or scaled up by a factor of ten. Previously, we investigated the perception of physics in VR when subjects were scaled down by a factor of ten (\citealt{pouke2020plausibility}). We found out that subjects considered a physics model of regular human scale to be more realistic than an accurate approximation of physics in the scaled down environment. This offered additional proof of humans being oriented to Newtonian physics taking place at human scale, and anything deviating much from that scale appears unnatural. In this paper we extend our prior work by considering scaled-up subjects allowing us to compare the perception of physics in VR both in small and large scales.



Currently, not much is known about how scaling a person would affect their perception of physical phenomena, such as accelerations. Interestingly, if we consider the interaction of scaled-down characters with their surroundings in many works of fiction, the tendency to represent the world as scaled up in comparison to normal-sized protagonists can be observed. Early examples can be seen in the classic film \textit{The Incredible Shrinking Man}. When the main character throws grains of sand off the table while insect-sized, the grains accelerate and fall as if they were boulders - when they should be falling down instantly. Similarly, when the character is awash with rainwater holding onto a pencil, the water and the pencil act more akin to a river and a log when the pencil should be bobbing with few waves and no visible whitewater should be apparent. Although the deficiencies in the realism of the \textit{Incredible Shrinking Man} can be attributed to 1950s technologies, similar inaccuracies still remain in modern movies from \textit{Honey I Shrunk the Kids} to \textit{Downsizing}. These inaccuracies are not necessarily resulting from directors’ lack of understanding of physics, but might be conscious choices to represent what the viewers would expect. 

\begin{figure}[t]
\begin{center}
\includegraphics[width=0.5\textwidth]{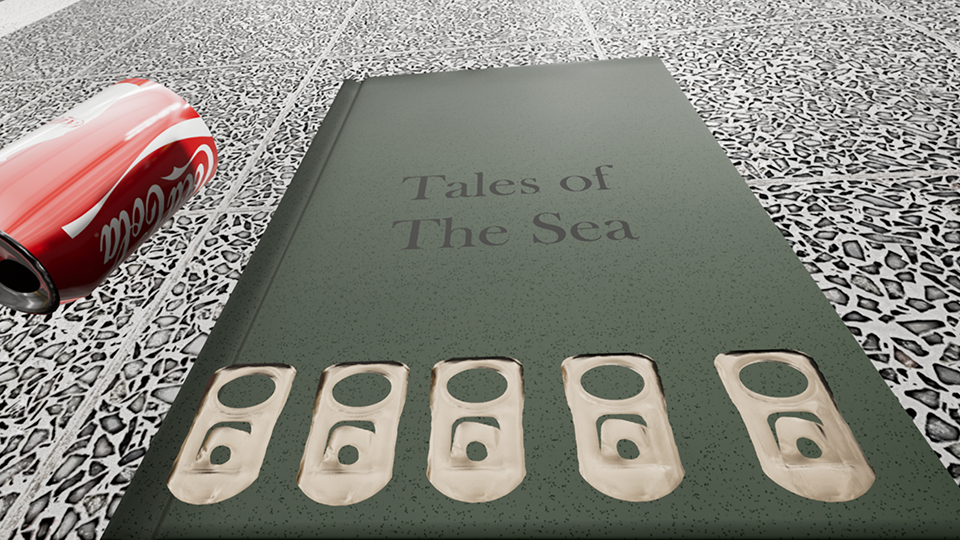}
\end{center}
\caption{First person perspective from the subject's point of view in the VE at the start of the experiment in Study 1.}
\label{book}
\end{figure}

VR and telepresence applications allow humans to live through experiences such as the \textit{Incredible Shrinking Man} through the eyes of a scaled-down entity. A specific category of virtual environments (VEs) providing such experiences are multiscale collaborative virtual environments (mCVEs), in which multiple users can collaborate in, for example, architectural or medical visualizations across multiple, nested levels of scale (e.g., \cite{kopper_design_2006, zhang_mcves:_2005}). In addition, the scaling of users has been utilized in several collaborative mixed reality (MR) systems (e.g., \cite{billinghurst_magicbook:_2001, piumsomboon2018mini, piumsomboon_superman_2018}). Teleoperation of robots can allow humans to interact with the physical world at micro- and nanoscale. Similar to mCVEs, robotic teleoperation systems using multiple scales are beginning to emerge (\citealt{izumihara2019transfantome}). Although teleoperation in the physical world can leverage stereoscopic camera systems resembling immersive VR applications (\citealt{hatamura_direct_1990}), purely virtual representations leveraging computer graphics can be used in, for example, educational and training systems for micro- and nanoscale tasks (\citealt{bolopion_review_2013, millet_improving_2008}). Robotic surgery systems can perform operations at a microscopic level (\citealt{hongo_neurobot:_2002}) whereas stereoscopic VR can be utilized in telesurgery (\citealt{shenai_virtual_2014}). The benefits of VEs have been identified in various design and prototyping processes (\citealt{mujber_virtual_2004}) that can be extended into small-scale VEs, as well. Already two decades ago, both the design (\citealt{li_virtual_2001}) and assembly (\citealt{alex_virtual_1998}) of microelectromechanical systems (MEMS) were prototyped through desktop VEs. Recent studies have also investigated self-scaling as a method to help with aspects related to architectural and interior design (\citealt{zhang2020spatial, zhang2020exploring}).

Understanding human perception of scale-varying phenomena will be useful for the future design of applications such as those listed above. Although existing research has addressed many perceptual questions, such as the perception of distance and dimensions after altering one's virtual size (e.g., \cite{van_der_hoort_being_2011, banakou_illusory_2013, kim2017dwarf}), the perception of the behavior of physical objects has received relatively little attention. There are many potential future use cases for user scaling that might require interaction with physical or physically simulated objects. We argue that it is not intuitive for humans to correctly perceive physical phenomena, such as rigid body dynamics, in scales that differ greatly from a normal human scale. An object dropped from 20 cm takes significantly less time to fall than an object dropped from 2 m, and their perceived accelerations are different. Additional physical phenomena, such as fluid dynamics, frictions, and static electricity might affect interactions even further as the scale of the operations becomes smaller. For this reason, additional consideration is required when designing systems in which real or virtual interactions take place on atypical scales, and thus it is important to understand human perception of physical phenomena on those atypical scales.

In this paper, we present our results on human perception of physics at abnormal scales. First, we focus on the mismatch between perceived realism and a physically accurate approximation of reality when interacting in a VE while scaled down by a factor of ten. Then, we present the results of a similar study where subjects were scaled up by a factor of ten and compare the results between the two studies. Based on previous research, we believe that humans generally perceive themselves at the correct scale when presented with mismatching size cues, as long as visual body cues are present (\cite{langbehn_scale_2016-1}). We also believe humans are generally accustomed to rigid body dynamics taking place at a human scale and under normal gravity conditions (\cite{mcintyre2001does}). Therefore, we hypothesize that subjects neither accept the realistic approximation of physics at an abnormal scale, nor are they blind to changes in scale. Instead, when presented with two different scale-dependent rigid body dynamics, they are more likely to consider the physically inaccurate one to be the more perceptually realistic one.

This paper is structured as follows. Section 2 reviews previous research related to this work. Section 3 presents the research method, experimental setup and the results of Study 1. Section 4 similarly reports Study 2 and also compares the results of both studies. Section 5 discusses our findings and Section 6 concludes the paper.

\section{Related Research}
\subsection{Perspective}
The manipulation of a user's scale can be accomplished by changing various properties of the virtual character the user is controlling in the VE. Changing these properties has various subjective effects. When scaling a user's virtual size, one of the most obvious properties to change is the viewpoint height, as it defines the virtual camera origin in relation to the VE, simulating a change in physical size. Viewpoint height affects egocentric distance perception (\citealt{leyrer_influence_2011, zhang_mcves:_2005}). Interestingly, minor changes in viewpoint height might go unnoticed by users (\citealt{leyrer_influence_2011,deng2019floating}). Users' interaction capabilities such as locomotion speed and interaction distance can be changed according to scale, depending on the purpose of the application (\citealt{zhang_mcves:_2005}). When using a head mounted display (HMD), the scaling of the user can also affect the virtual interpupillary distance (IPD), which is the distance between the two virtual cameras that are used to render the environment for the user. Changing this distance can affect the user’s sense of their own size relative to the VE (\cite{piumsomboon_superman_2018, kim2017dwarf}).

\subsection{Body Scaling}
As already mentioned, \textit{body scaling} refers to humans utilizing their own body as a primary scale cue, hence the virtual representation of the user's body greatly affects their perception of sizes and distances in a VE (\citealt{8798040, ogawa2017distortion}). \cite{linkenauger_welcome_2013} studied the role of one's hand as a metric for size perception; they conducted an experiment where they scaled the users’ virtual hand and found out that it had a strong correlation with perceived object size. \cite{8798040} studied the effect of hand visual fidelity on object size perception and found that the visual realism of the hand affects the extent of the body scaling effect. \cite{van_der_hoort_being_2011} embodied the entire user in a doll’s body as well as in a giant’s body using a stereoscopic video camera system and an HMD. They found that the embodiment significantly affected the users’ distance and size perceptions, especially if the user experienced a strong body ownership illusion (\citealt{slater2009inducing}) with the virtual body. \cite{banakou_illusory_2013} compared the effects of embodying the user as a child versus as a scaled-down adult. They found that the effect of altered size and distance perceptions was even larger when embodied as a child, and it also made the users associate themselves with childlike personality traits. 

\subsection{Environmental Cues}
The environment, whether real or virtual, affects the perception of scale. There is evidence of humans generally underestimating egocentric distances in VEs, except when the VE is faithfully modeled to represent a real environment (\citealt{renner_perception_2013}). However, if a familiar room is scaled slightly up or down, underestimations are reintroduced (\citealt{thompson_elucidating_2007}). Familiar size cues also affect the sensitivity to eye height manipulations (\citealt{deng2019floating}). \cite{langbehn_scale_2016} studied the effect of body and environment representations as well as the scale of external avatars on users' perception of dominant scale in mCVEs (the dominant referring to the “true” scale in an mCVE system where users can coexist in multiple scales). They found that humans tended to use their body as the primary metric for judging their own size and the environment if the representation of one's own body was not available. In addition, an environment with familiar size cues helps in the determination of scale, whereas an abstract environment does not. They also found that the majority of subjects tended to estimate external avatars to be at the dominant scale instead of themselves.

\begin{figure}[t]
\begin{center}
\includegraphics[width=1\textwidth]{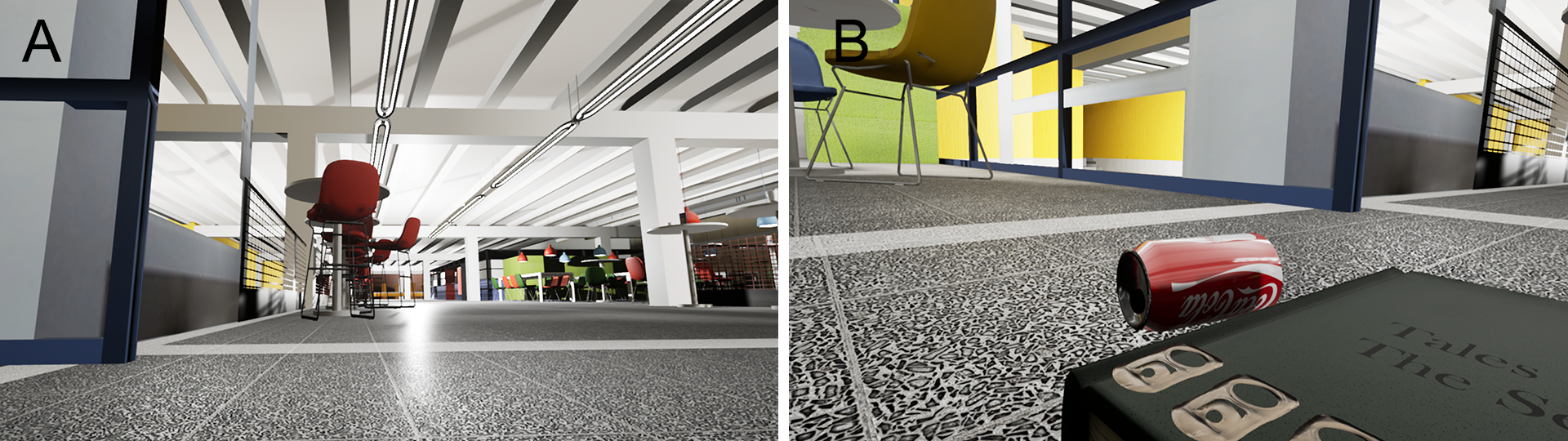}
\end{center}
\caption{A screenshot of the VE from the subject's perspective when looking forward and upward with book and tabs below the line of sight \textbf{(A)} and when looking left \textbf{(B)}.}
\label{scene1}
\end{figure}

\subsection{Perception of Physics}
Previous research suggests that humans have an internal physics model according to which they expect the world to function. Studies in micro- and nanoscale teleoperation have revealed that, due to changes in physics, interactions at these scales can become difficult for the human operator, but education inside virtual reality environments has been found to alleviate this drawback (\citealt{millet_improving_2008, sitti_microscale_2007}). \cite{mcintyre2001does} reported a study in which astronauts' movements to catch a vertically moving ball were more inaccurate in zero gravity (0g) in comparison to earth gravity (1g); this was interpreted as evidence that the central nervous system utilizes an internal model of gravity in addition to visual judgement of acceleration to synchronize movements. \cite{senot2005anticipating} used VR to study human estimation capabilities to intercept moving balls and found further evidence on subjects being more capable of intercepting objects accelerating according to normal gravity. \cite{yao2006experiment} created a haptic illusion of an object rolling or sliding inside a cavity and studied the subjects' capability of estimating the lengths of virtual tubes. According to their results, the subjects performed better than chance in estimating the tube lengths even using reduced sensory cues, indicating a capability to estimate object movements under the influence of gravity. 

\cite{ullman2017mind} compared humans' internal physics model to a contemporary game engine. Their findings suggest that although humans are not entirely capable of accurately predicting object motions, they are capable of making noisy, "good enough" approximations of Newtonian physics which can be compared to the physics simulations generated by physics engines that are integrated in contemporary game engines. \cite{mccoy2019judgments} asked more than a thousand subjects to rate the 'effort' required by various imaginary magical spells violating physics and found the subjects' responses as strikingly consistent. Despite describing completely imaginary phenomena, the subjects were very consistent in defining relative efforts that seemed to depend not only the type of the spell (such as levitate or conjure) but the size of the target as well. Although this finding is not directly related to the perception of physical phenomena, it again speaks for internal intuitive physics model that is consistent across humans.

\subsection{Presence and Plausibility}
The concepts of immersion (\citealt{slater_framework_1997}), presence and plausibility (\citealt{slater_place_2009}) are relevant for this study. In Slater’s classical definition, the level of immersion refers to the level of technical fidelity of the VR system (i.e., resolution, field of view, vividness of graphics; \citealt{slater_framework_1997}). In addition, the realism of the user’s response to the VR system depends on two orthogonal components, presence or place illusion (PI) and the plausibility illusion (PSI; \citealt{slater_place_2009}). PI refers to the sensation of being in another place, whereas PSI refers to the perceived credibility of the virtual scenario or experience (the illusion of being there versus the realness of what is happening; \citealt{rovira_use_2009}). PSI depends on the extent to which the VE can produce authentic responses for user actions. \cite{rovira_use_2009} argued that for PSI to occur, participants must perceive themselves as beings that exist in the VE; user actions must elicit actions in the VE and the VE must acknowledge the user (for example, virtual characters react to the user). In addition, the VE should match the users’ prior knowledge and expectations. \cite{skarbez_psychophysical_2017} used the term \textit{coherence} to refer to the aspects of a VE that contribute to PSI, such as virtual humans and the behavior of virtual objects. They argued that although immersion is a technical attribute that affects PI, coherence is a similar technical attribute affecting PSI.

In Study 1, we used the concept of PSI to study human perception of the behavior of physical objects while the subject was scaled down and interacting in a normal-sized environment. In Study 2, we repeated the same procedure for scaled-up subjects. However, we delimited virtual characters out from the scope of in both studies. Instead, we were interested in how subjects would perceive the coherence in terms of behavior of virtual objects, when it would be reasonable to expect a mismatch between expectations and correctly simulated reality. In addition, we investigated whether the extent of PI affected PSI in this particular context.

Building on the terminology discussed by \cite{skarbez2017survey, skarbez2020immersion}, the phenomenon studied in this paper could also be referred to as \textit{coherence} - \textit{fidelity} mismatch; the logic expected by the users mismatches with more faithful representation of reality. It is expected that coherence differs from reality in, for example, fantasy games or other entertainment applications where PSI is maintained even when unearthly phenomena are taking place. However, we consider the mismatch studied here to be specifically interesting due to its implications for VR and telepresence applications taking place at abnormal scales.

\section{Study 1}

\subsection{Study Design}
\subsubsection{Physics Conditions}
The specific objective of Study 1 was to investigate the PSI of subjects in two different physics conditions. The purpose of both conditions was to visually represent a scaled-down subject in a normal-sized environment, and the physics simulations differed between the conditions as follows. In the \textit{true physics} condition, the rigid body dynamics affect virtual objects in an approximately similar way to what would be accurate at that scale. In the \textit{movie physics} condition (named after physical behavior as typically seen in Hollywood movies in scenes depicting scaled-down characters), rigid body dynamics behave in what would be the approximation of a normal human scale. 

Our assumption was that the users would be able to distinguish the difference between \textit{true physics} and \textit{movie physics}, and we predicted that subjects would be more likely to expect and feel the \textit{movie physics} condition to be the more perceptually realistic representation. This would suggest a Plausibility Paradox, a mismatch between perceived realism and the correct approximation of realism. 

\subsubsection{Hypotheses}
We hypothesized that in the \textit{true physics} conditions, the behavior of physical objects would feel incorrect for subjects despite their knowledge of being virtually shrunk down. More specifically, our hypotheses were as follows: 

\begin{itemize}
\item [H1:] For a scaled-down user, \textit{movie physics} is more likely to feel realistic than \textit{true physics}.

\item [H2:] For a scaled-down user, \textit{movie physics} is more likely to match the user's expectations than \textit{true physics}.


\end{itemize}

\subsubsection{Virtual Environment}
We designed a VE for the two physics conditions described above using Unreal Engine 4.22 (UE). In both conditions, the scaling operations took place in one order of magnitude, giving the impression of a doll-sized perspective. We did not use full body tracking or attempt to induce a strong body ownership illusion (\citealt{slater2009inducing}), so there was no visualization of any body parts in the VE other than the subject's hands. We used the default UE VR hand visualization for interaction and to present a medium-fidelity body size cue (\citealt{8798040}). There was no difference between the conditions regarding how the hands functioned or how the user was able to move.  

To help in providing accurate size cues, we modeled the VE to resemble a location in the main corridor of the campus in which the study took place. The dimensions and materials of the VE were modeled after the real environment. In addition, we took measurements of various real-world objects, such as chairs, tables, and leaflets, which we modeled and scaled accordingly and placed in the VE as static objects.

The scaling of the user in the \textit{true physics} condition was achieved by shrinking the user with the UE's built-in \textit{World to Meters} parameter, which automatically scales the player character's height, virtual IPD and interaction distance. The skeletal meshes representing the player character's virtual hands were scaled down manually. In the \textit{movie physics} condition, the player character properties were kept as default and the VE was scaled up instead. The purpose for this approach was to give the visual illusion of a scaled-down user, while retaining physics conditions that correspond to the normal human scale. The sizes and relative distances of scene objects were increased by a factor of ten. In addition, the properties of lights and reflection capture objects were adjusted so that the overall visual appearance of both conditions were kept as similar as possible.

\subsubsection{Interaction Task}
The interaction task consisted of the manipulation of virtual soda can pull tabs approximately 3 cm in length and 1.9 cm in width (as presented in Fig. \ref{book}). The tabs were chosen for the experiment both for their small, consistent mass as well as for being a reasonably authentic object that could be seen in the simulated VE. We considered a lightweight object to be most practical for simulating throwing in VR so that we would not have to simulate the decrease in hand acceleration due to increased inertia at the end of the arm or limitations due to arm strength (\citealt{cross2004physics}). In both conditions, the subjects would try dropping and throwing five tabs. Picking up and throwing the tabs took place utilizing the default mechanism in UE, similar to contemporary VR applications in general. The subjects simulated grabbing objects by squeezing the trigger of the motion controller and dropping them by releasing the trigger. Virtual throwing took place by swinging the motion controller and then releasing the trigger, and the object thrown retained its velocity at the moment of release, simulating throwing.

In the \textit{true physics} condition, the tabs would drop down fast, similarly as to if they were dropped from the height of 15-20 cm (simulated falling speed approximately 0.175 s at 20 cm in UE). In addition, the throwing distances would appear short because of the limited velocity that can be actuated due to real hand movements scaled down by an order of magnitude. The \textit{movie physics} condition, on the other hand, simulated the tabs as falling down more slowly, similarly to an object dropped from human height (simulated falling speed approximately 0.6375 s at 2 m in UE). In addition, the throwing distances were much larger in the \textit{movie physics} condition due to the larger velocity that the subjects were able to actuate on the tabs by virtual throwing. 

Due to the simulated size, the tabs were also different between conditions in terms of their bounciness (there were no changes in physics simulation properties, such as restitution). In the \textit{movie physics} condition, the tabs bounced visibly off surfaces, or jittered slightly after being dropped. However, in the \textit{true physics} condition, there was little to no visible bounciness.  

The tabs were placed on top of a large book so that the subjects would not have to pick them up from the floor. The book also provided an additional size cue. We gave the book a neutral, non-distracting appearance and a general title so that it was recognizable as a book, but would not otherwise draw too much attention. A Coca-Cola can was placed as a familiar sized cue on the left side of the book. Fig. \ref{book} shows the book and the tabs as seen in the beginning of the simulation. Fig. \ref{scene1} A and B show the scene as seen at the beginning of the simulation when looking forward (A) and left (B).

The virtual mass of the tabs was set at 1g in both conditions. Default physics settings in UE were utilized, with the exception of turning on the physics sub-stepping for additional physics accuracy by enabling physics engine updates between frames. Drag by air resistance was set to zero in both conditions. The simulation itself ran at stable 80 FPS which is the maximum frame rate of Oculus Rift S.

\subsubsection{Participants}
The experiment was carried out as a within-subjects experiment, in which 44 subjects (23 females and 21 males) performed both conditions during one session. Two participants were excluded due to issues with the functionality of the VR equipment or due to vision impairments.  The order of the conditions was counterbalanced so that there was an equal number of male and female participants starting with each condition. The subjects' ages ranged from 19 to 66, mean and median ages being 30 and 26, respectively. The standard deviation for the ages was 10.4. The study was conducted either in English (12 females and 7 males) or in Finnish (11 females and 14 males), depending on the preference of the subject. Each participant was rewarded with a gift voucher of two euros.

\subsubsection{Experimental Procedure}
The experiment was set in a laboratory in which the subjects used the Oculus Rift S system with provided Oculus Touch controllers for the experiment. The Rift S has a variable IPD software setting, so the IPD was set to 62.5 for females and 64.5 for males, the closest approximation available based on the averages reported for adults by \cite{dodgson2004variation}. In the beginning of a session, the subject read through a written \textit{Information for Subjects} document and signed an informed consent sheet. The subject was then instructed on using the VR hardware, specifically how to use the Rift S Touch motion controllers for picking up and throwing objects. Next, the subject was instructed to stand on a particular starting spot in the laboratory marked with a masking tape. When the subject was wearing the HMD and the motion controllers comfortably, an instruction script was read in English or Finnish. The script stated that the subjects were at the university central hallway, shrunk down 10-fold to a size of a doll, and were to drop and throw the tabs placed on top of a book in front of them.



Active noise-cancelling headphones were placed on the subject to block out any potential external noise from other rooms in the building, and then the experiment began. After performing both conditions, the headphones and the VR hardware were removed and the subject was asked to respond to a post-experiment questionnaire as well as a background questionnaire on a different laptop. The subject was asked for any additional comments or questions, and if he/she could be contacted for future studies, and then given her/his gift voucher. The average duration of the session was 20 minutes per subject.


\subsubsection{Questionnaires}
We collected plausibility related data using two forced choice questions (main questions 1 and 2), two open-ended questions (O1 and O2) and a 7-point  Likert scale questionnaire regarding the behavior of the tabs (L1-L5). In addition, the subjects filled out the extended version of the Slater-Usoh-Steed (SUS) Presence questionnaire (\citealt{slater_depth_1994, usoh2000using}), as well as a background information questionnaire. The main questions 1 and 2 were as follows:    

\begin{enumerate}
    \item \textit{Thinking back how the pull tabs were behaving in the experiment, which felt more realistic (like what would happen in the real world if you had been shrunk down), the first or the second time? }
    \item \textit{Thinking back how the pull tabs were behaving in the experiment, which matched your expectations (similar to what would happen in the real world if you had been shrunk down), the first or the second time?}

\end{enumerate}
The main questions were coupled with open-ended questions (O1 and O2), that were simply stated as \textit{"Why?"}. The purpose of the open-ended questions was to evaluate to what extent the subjects' responses were related to the physics or other reasons.

The forced-choice and open-ended questions were followed by a 7-point Likert scale questionnaire asking subjects to judge how they perceived various aspects related to the behavior of the tabs. Each question was stated twice in the questionnaire, referring to the first time and the second time subject interacted with the tabs (either using the \textit{true physics} and then the \textit{movie physics} or vice versa). The first three questions (L1-L3) were bipolar, whereas the last two (L4, L5) were unipolar. The Likert questions L1-L5 and their associated scales were as follows:  

\begin{itemize}
    \item [L1] \textit{The falling speed of pull tabs (too slow, too fast)}
    \item [L2] \textit{The speed of pull tabs when thrown (too slow, too fast)}
    \item [L3] \textit{The distance of pull tabs when thrown (too close, too far)}
    \item [L4] \textit{The way the pull tabs were bouncing when thrown (incorrect, correct)}
    \item [L5] \textit{The impact of gravity on the pull tabs (incorrect, correct)}
\end{itemize}

Similarly to Study 1, we gathered qualitative data, subject background data as well as questionnaire data to better understand the responses given by the subjects.

All questions were presented in either English or Finnish, depending on which was chosen as the preferred language by the subject when signing up for the experiment.


\subsection{Results}
\subsubsection{Hypotheses}
According to the responses to the main questions, the majority of the subjects considered the \textit{movie physics} condition as the more realistic one. Out of 44 subjects, 32 participants (73\%) responded to the first question that they considered the \textit{movie physics} condition more realistic, which confirms H1. For the second question, 40 out of 44 (91\%) subjects responded that the \textit{movie physics} matched their expectations better, which confirms H2. Furthermore, we analyzed the frequencies of responses to questions 1 and 2 with a binomial test and found their corresponding two-tailed p values as $p=0.004$  and $p = 1.7051^{-8}$, respectively. From this we can conclude that it is unlikely that the responses to questions 1 and 2 were due to chance. In addition, this indicates that subjects were able to distinguish between the two physics conditions and more consistently selected the \textit{movie physics} condition, which was the inaccurate physics condition.

Out of twelve respondents who considered \textit{true physics} more realistic, nine responded that the \textit{movie physics} matched their expectations more. Only one subject considered the \textit{movie physics} more realistic while simultaneously stating that the \textit{true physics} better matched her/his expectations. 

\subsubsection{Understanding Contributing Factors}
We gathered supplementary data to further understand the results. These data include responses to open-ended questions O1 and O2, Likert-scale questions L1-L5, as well as subject background and self-reported level of presence.

\begin{figure}[t]
\begin{center}
\includegraphics[width=1\textwidth]{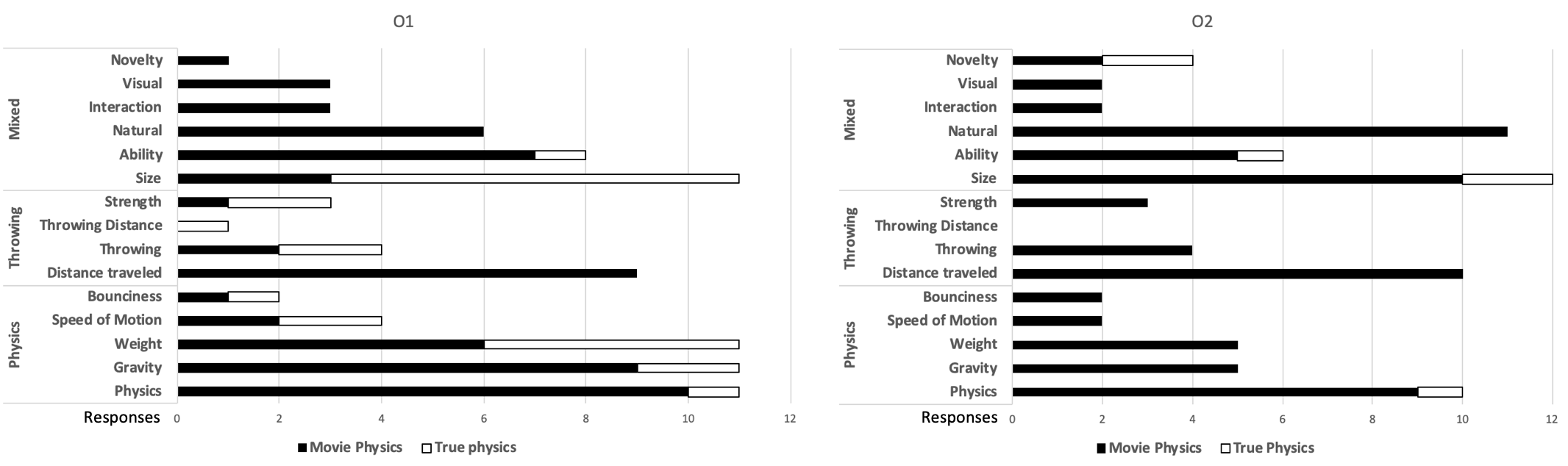}
\end{center}
\caption{Qualitative codes for responses O1 (\textbf{A}) and O2  (\textbf{B}) according to perceived realism in Study 1.}
\label{VRIISqstack2}
\end{figure}

The purpose of the open-ended questions was to evaluate to what extent the subjects' responses to the main questions 1 and 2 were related to the perceived realism of the physics. The responses consisted of one-sentence statements typed by the subjects. Thematic analysis with an inductive approach (e.g., \cite{patton2005qualitative}) was carried out independently by two researchers and used to identify codes in the response data. A summary of the codes can be viewed in Fig. \ref{VRIISqstack2} A and B). Examples of responses in O1 can be seen in Table \ref{S1O1examples}, whereas examples of responses in O2 can be seen in Table \ref{S1O2examples}.



\begin{table}[]
\tiny
\centering
\caption{Examples of O1 responses (justification to main question 1) in Study 1.}
\begin{tabular}{|p{3cm}|p{3cm}|p{10cm}|}
\hline
\textbf {Codes} & \textbf {Preference} & \textbf{Response} \\
\hline
\hline
\textit{Gravity, natural} & \textit{\textit{Movie physics}} & \textit{"Gravity felt more natural"}\\
\hline
\textit{Gravity, natural} & \textit{Movie physics} & \textit{"At the second time, the objects fell to the ground faster, which felt unreal"}\\
\hline
\textit{Gravity} & \textit{True physics} & \textit{"I think when the height of the object is not that high, it should reach the ground faster."}\\
\hline
\textit{Physics, visual} & \textit{Movie physics} & \textit{"movement in space felt more realistic, but the objects lacked 3D, ring pulls are not paper thin"}\\
\hline
\textit{Ability, distance traveled, physics} & \textit{Movie physics} & \textit{"because I was more comfortable with the controllers after using them for some time, and i knew i could do more things now like throwing more far away after some time, and also they were moving more smoothly"}\\
\hline
\textit{Bounciness, throwing, distance traveled, ability} & \textit{Movie physics} & \textit{"I am not sure but I think the second time they still moved a bit after I dropped them to the floor, before being completely still. I think I also managed to throw one of the pull tabs the second time, which felt more realistic than them dropping very quickly just right in front of me after I tried to throw them (but this could also just have been my inability to throw the first time)."}\\
\hline

\textit{Weight} & \textit{Movie physics} & \textit{"Second time they felt too heavy".}\\

\hline
\textit{Weight, strength, size} & \textit{True physics} & \textit{"Pull tabs are not heavy and when I'm small, I probably would not have the strength to throw them afar".}\\
\hline

\textit{Ability} & \textit{Movie physics} & \textit{"I was able to act more normal in the second round. I had worked out the mechanics of the VR better and spent less time attempting to make the task work".}\\
\hline

\end{tabular}
\label{S1O1examples}
\end{table}

\begin{table}[]
\tiny
\centering
\caption{Examples of O2 responses (justification to main question 2) in Study 1.}
\begin{tabular}{|p{3cm}|p{3cm}|p{10cm}|}
\hline
\textbf {Codes} & \textbf {Preference} & \textbf{Response} \\
\hline
\hline
\textit{Distance traveled} & \textit{Movie physics} & \textit{"As I was taking a swing with my arms I was expecting them to land far away from me which they did only during the first time."}\\
\hline
\textit{Speed of motion} & \textit{Movie physics} & \textit{"In the second time the tabs were falling down surprisingly fast"}\\
\hline
\textit{Size, weight} & \textit{Movie physics  (different from O1)} & \textit{"I was not thinking I was shrunk. So it felt estrange to have such heavy pull tabs"}\\
\hline
\textit{Physics, size} & \textit{Movie physics (different from O1)} & \textit{"I didn't think at first (until I saw the previous question) shrinking down would also affect the time it takes for the objects to reach the ground. The physics first time behaved just like in normal life."}\\
\hline
\textit{Size} & \textit{Movie physics (different from O1)} & \textit{"Even though I knew I was shrunk down, I still could not think that way when doing the experiment"}\\
\hline
\textit{Natural, physics} & \textit{Movie physics} & \textit{"The behavior seemed more natural, although probably the laws of the physics tell otherwise"}\\
\hline
\textit{Ability, throwing} & \textit{Movie physics (different from O1)} & \textit{"I thought throwing the pull tabs would be relatively easy, like in the second time".}\\

\hline
\textit{Weight} & \textit{Movie physics} & \textit{"Intuitively I figured things would be light".}\\
\hline

\textit{Size, novelty} & \textit{True physics} & \textit{"I felt that I was really small in that world for the first time.".}\\
\hline
\textit{Natural} & \textit{Movie physics} & \textit{"First time. Felt somehow more natural. They didn't have much difference, though".}\\
\hline
\end{tabular}
\label{S1O2examples}
\end{table}

In short, the responses to questions O1 and O2 indicate that majority of users (38 out of 44) made their choices primarily according to reasons related to the behavior of the physically simulated tabs. Other primary reasons were related to general interaction and becoming accustomed to controllers. Few references were made to visual details (appearance of tabs and colors) as secondary reasons or general remarks. 

\subsubsection{Likert Data}
\begin{figure}[t]
\centering
\includegraphics[width=1.0\textwidth]{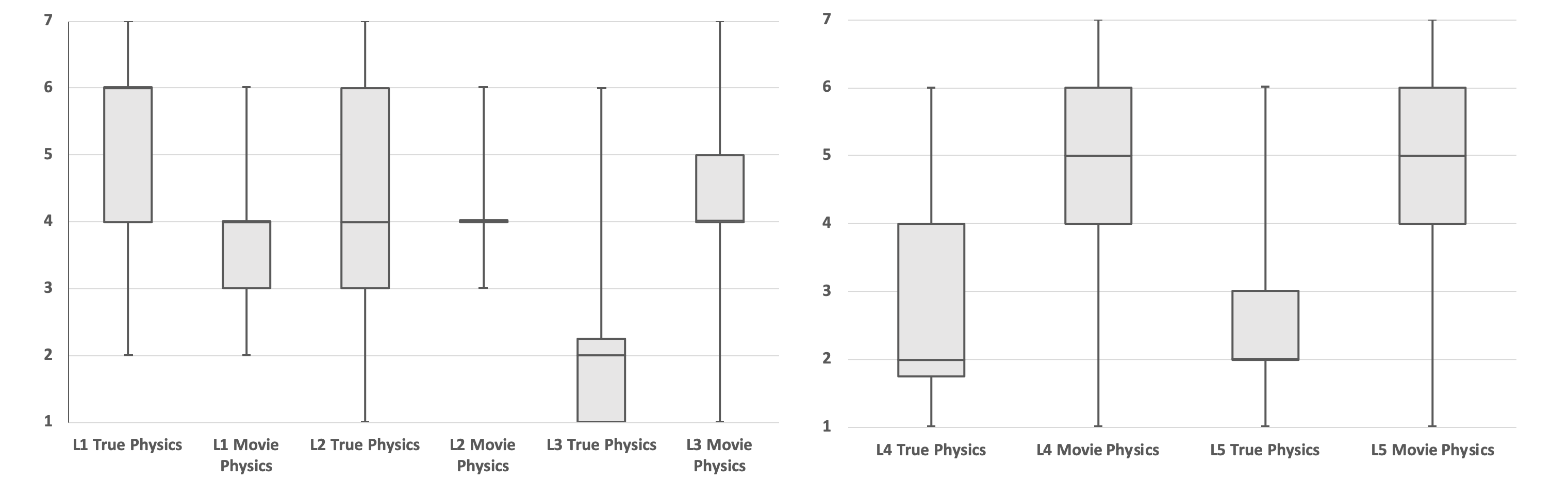}
\caption{Study 1 Likert responses L1-L3 (\textbf{A}) and L4-L5 (\textbf{B}) visualized as box plots, interquartile ranges and minimum and maximum responses. In (\textbf{A}), responses closer to 4 are perceived as closer to realism, whereas in (\textbf{B}) responses closer to 7 are perceived as closer to realism.}
\label{Likert13}
\end{figure}

Inspecting the Likert responses for questions L1-L5, we found that the \textit{movie physics} condition was closer to perceived realism (median responses closer to 4 in questions 1 and 3 and closer to 7 in questions 4 and 5) in all questions except L2, in which the median response was the same for both conditions. We analyzed the responses to questions L1-L5 with the Wilcoxon Signed Rank test and found that the responses were significantly different (p \textless 0.05) for all questions except L2 (p = 0.845). This gives additional confirmation that the subjects perceived the \textit{movie physics} condition more realistic due to differences in the behavior of the physically simulated tabs. A summary of responses including, median, mode and standard deviation for questions L1-L5 can be seen in Table \ref{Likertdata}. In addition, box plots visualizing the medians, interquartile ranges as well as minimum and maximum responses can be seen in Fig. \ref{Likert13} A and B.

\begin{table}[]
\small
\centering
\caption{Summary of Likert data in Study 1. Responses perceived closer to realism are emphasized in bold.}
\begin{tabular}{|l|l|c|c|c|}
\hline
\textbf {Question} & \textbf {Condition} & \textbf{Median} & \textbf{Mode} & \textbf{STD} \\
\hline
\hline
\textbf{L1:} & \textit{true physics} & 6 & 6 & 1.4 \\
Falling speed & \textit{movie physics} & \textbf{4} & \textbf{4} & 0.8         \\
\hline
\textbf{L2:} &  \textit{true physics} & \textbf{4} & 6 & 1.8 \\
Speed when thrown & \textit{movie physics} & \textbf{4} & \textbf{4} & 0.9 \\
\hline
\textbf{L3:} &  \textit{true physics}  & 2 & 2 & 1.3 \\
Distance when thrown & \textit{movie physics}  & \textbf{4} & \textbf{4} & 1.1 \\
\hline
\textbf{L4:} & \textit{true physics}  & 2 & 2 & 1.6 \\
Bounciness & \textit{movie physics}  & \textbf{5} & \textbf{6} & 1.5 \\
\hline
\textbf{L5:} & \textit{true physics} & 2 & 2 & 1.4 \\
Gravity & \textit{movie physics} & \textbf{5} & \textbf{6} & 1.4 \\
\hline
\end{tabular}
\label{Likertdata}
\end{table}

\subsubsection{Effect of Background and SUS scores}
Furthermore, we used a binary logistic regression to analyze the effects of subject background and presence on their responses to main question 1. We used \textit{educational background, gender, age, vr experience, gaming experience, SUS average and SUS score} as independent variables and the response to main question 1 as the dependent variable. 

For analysis purposes, we transformed the Background Questionnaire responses to \textit{educational background } into a binary variable consisting of roughly equal sized groups of \textit{natural sciences and engineering} (25 subjects) and \textit{social sciences} (19 subjects). In addition, the open responses to \textit{VR experience} and \textit{gaming experience} was transformed into respective ordinal variables ranging from 0 (no experience) to 4 (plenty of experience). When interpreting the \textit{gaming experience} responses, additional emphasis was given to recent experience as well as experience regarding PC and console based 3D gaming (such as first person shooters and simulators) due to the tendency of such games to contain game physics simulations similar to those used in this experiment. The responses to SUS scores were transformed into two ordinal variables consisting of the average of responses as well as the computed SUS score. 

The logistic regression model was unable to predict the response using the independent variables. The model explained 17\% of the variance (Nagelkerke's $R^2$) in perceived realism. Although the overall classification rate was 72.7\%, only 16.7\% (two responses) of the \textit{true physics} responses were correctly classified. None of the independent variables had a significant effect on the prediction of the response (p = 0.184 - 0.858). According to this analysis, the perception of realism was not significantly affected by the background, education or gaming experience of our subjects. The level of presence according to self-reported SUS score did not have any effect either.

\subsubsection{Perception of Mass and Strength}
Although we never queried subjects directly regarding the physical properties of the tabs themselves, several subjects commented on the weight of the tabs or their own strength when interacting with the tabs. Five of the subjects who responded in English commented on the feeling of the perceived heaviness of the tabs (see Table. \ref{S1O1examples}). It is interesting to consider these spontaneous responses regarding differences in the weight of the tabs given than there was no change in the controllers that the subjects used for each condition. This could be an indication of a pseudohaptic effect (\citealt{lecuyer2009simulating}) (for example, manipulating the control-to-display ratio of the visual feedback when lifting an object can give the user an illusion of increased weight (\citealt{samad2019pseudo})). However, it is possible that the subjects were simply referring to the visible trajectories and falling speed of objects (as in the tabs \textit{seemed} heavier instead of the tabs \textit{feeling heavier}). Several of the responses in Finnish specifically contemplated the assumed weight of the tabs in regards to how more much power they would have needed to use to throw the tabs given their reduction in size. To investigate these findings further, we added additional pseudohaptic related questions in Study 2.




\section{Study 2}
\subsection{Study Design}
In Study 2, we wanted to investigate the perception of rigid body dynamics while the user was enlarged by a factor of ten. We followed a methodology similar to Study 1 so that we could easily compare subjects' perceptions in small and large scales. We introduced minor methodological changes described below.

\subsubsection{Hypotheses}
Our hypotheses for Study 2 were similar to those of Study 1.


\begin{itemize}

\item [H3:] For a scaled-up user, \textit{movie physics} is more likely to feel realistic than \textit{true physics}.

\item [H4:] For a scaled-up user, \textit{movie physics} is more likely to match a user's expectations than \textit{true physics}.


\end{itemize}

\subsubsection{Virtual Environment}
\begin{figure}[h!]
\begin{center}
\includegraphics[width=17.5cm]{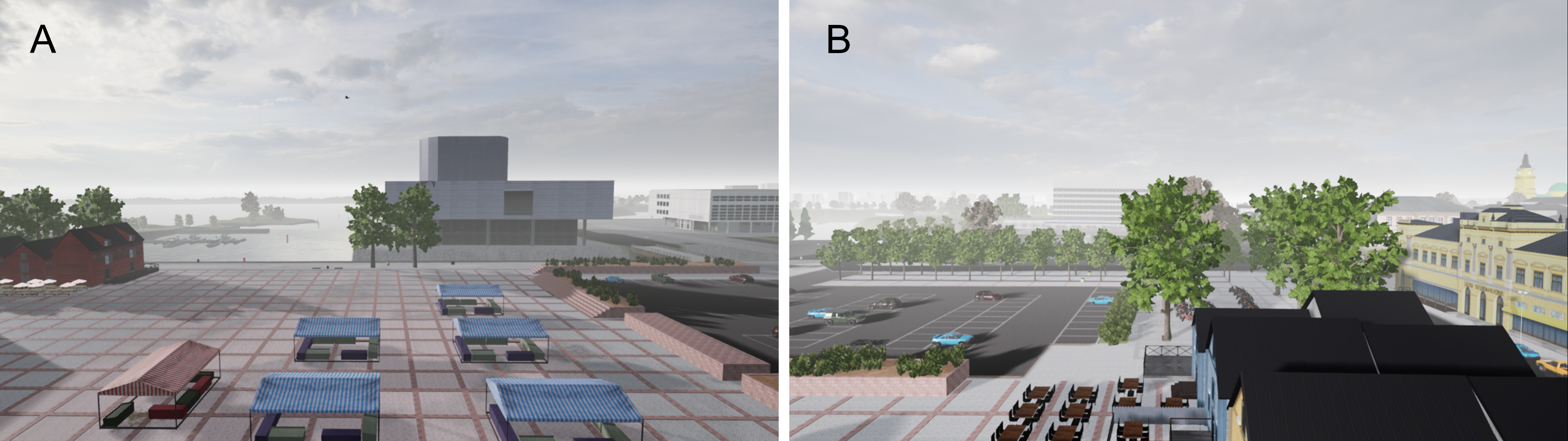}
\end{center}
\caption{The VE used in Study 2 as seen by subjects in the beginning of the experiment when looking forward \textbf{(A)} and right \textbf{(B)}}\label{fig:1}
\end{figure}

Study 2 portrayed the subject as a giant, 10 times larger than a regular human. Similarly to Study 1, the VE was also based on a real-world environment we expected to be familiar for most of our subjects. More specifically, the VE depicted a marketplace and its surroundings located in the center of the City of Oulu, Finland. The environment used 3D assets from the "Virtual Oulu" model described in \cite{alatalo_virtualoulu:_2016}. The assets were imported into a UE 4.24 scene. Some of the original materials were remade to follow a contemporary physically-based rendering (PBR) workflow for improved aesthetics. To enrich the model with additional size cues, the marketplace area of the model was augmented with additional detail such as street furniture, trees and foliage that were placed using Google Maps photographs and satellite photos as reference. GIS data from the City of Oulu were used to generate non-textured faraway buildings seen in the background of the scene. Also, generic textured building blocks were used in some areas to generate buildings not present in the original Virtual Oulu model but close enough to the viewer so that untextured models were not feasible. Although our aim was to make the scene appear realistic for the subjects, we took minor liberties in the placement of certain scene objects to make the scene more appropriate for the experiment. Namely, the immediate marketplace surroundings were left relatively empty to prevent  the subjects from hitting random objects and making unwanted plausibility noise. In addition, the position of trees next to the shoreline were adjusted so that the logs have a free passage to water (see Fig. \ref{fig:1} A). 


In addition to Virtual Oulu assets, GIS data and self-modeled assets, several commercial packages from the Unreal Marketplace were used in the VE. Animated seagulls and pigeons from the \textit{Birds} package were scaled to correct size (approximated wingspans 70 cm and 140 cm, respectively) and deployed in the scene to provide animated size cues. The commercial packages \textit{"Nordic Harbour", "Country Side", "Vehicle Variety Pack", "Modern City Downtown", "Sky Pack" }as well as \textit{"Trucks and Trailers"} were also utilized for foliage, vehicles, street furniture and other minor details, such as traffic signs. Water shader and buoyancy for logs was generated using the \textit{Waterline Pro} package. Screenshots of the scene can be seen in Fig. \ref{fig:1} A and B. 

Similarly to Study 1, the scale-changing effect was achieved by scaling the world-to-meters parameter of UE and player character properties, this time making the user to appear 10 times larger instead of smaller. Similarly to Study 1, we defined the rigid body dynamics as simulated by the game engine to act as the \textit{true physics} condition. For the \textit{movie physics} condition, we upscaled the \textit{default gravity Z} and \textit{bounce threshold} (as instructed by UE when scaling gravity) properties by 10 to generate conditions similar to human scale. This approach was taken to avoid generating two different-sized versions of the level so that we could eliminate visual difference. These approaches resulted in object free fall times of 1.97 seconds and 0.68 seconds, when object is dropped from the height of 18m in \textit{true physics} and \textit{movie physics} respectively.


\subsubsection{Interaction Task}
\begin{figure}[h!]
\begin{center}
\includegraphics[width=17.5cm]{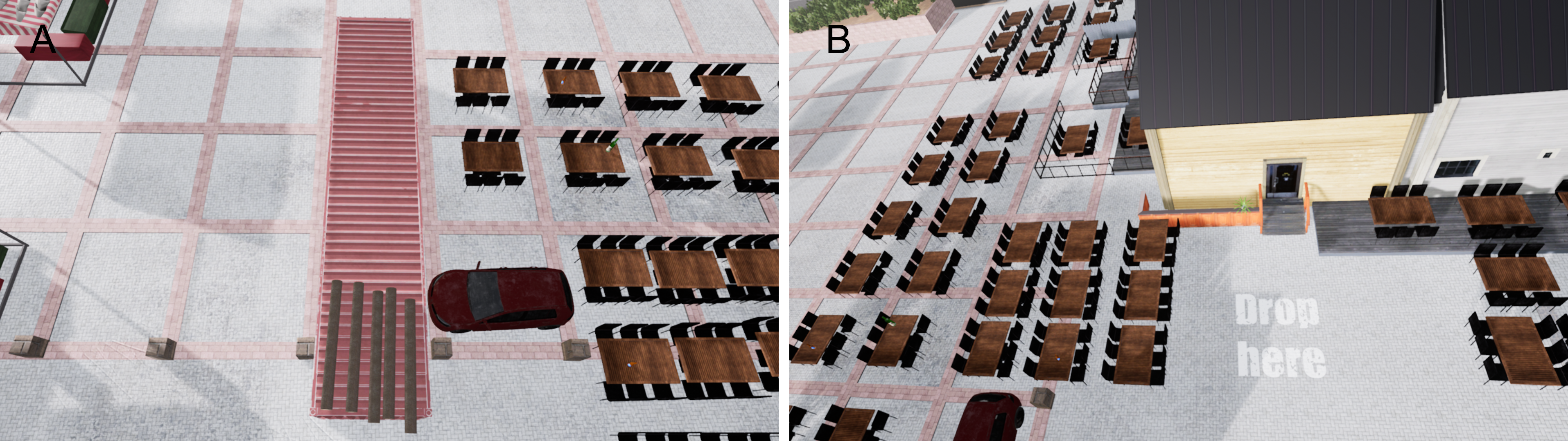}
\end{center}
\caption{The container with the logs as seen in the beginning of the experiment \textbf{(A)} and the area in which the first three logs were to be dropped \textbf{(B)}}\label{fig:2}
\end{figure}

The interaction task in Study 2 resembled the task in Study 1, consisting of dropping and throwing objects. Since the subject was a giant instead of doll-sized, the objects used in the interaction task were larger as well. Considering a handful of alternatives, we determined wooden logs as suitable objects for interaction, since they are somewhat familiar sized objects for most locals and frequently seen around town after being culled from local forests. The logs were approximately 2.9 m in length and 26.7 cm in diameter, matching the dimensions and mass of commercial pine logs. We placed the logs on top of a container, so that the subjects would not need to reach all the way down to the ground to grab the logs. The container with the logs can be seen in Fig. \ref{fig:2} A. 

In Study 1, the subjects were allowed to drop and throw the pull tabs in any way they wished. However, in Study 2 we instructed the subjects to drop exactly three logs to their right and throw two logs into the sea visible in front of the subjects (see Fig. \ref{fig:1} A). We also placed a "Drop Here" text on the ground to depict where exactly the logs should be dropped (see Fig. \ref{fig:2} B). There was a particle splashing effect when the logs hit the water surface as well a buoyancy effect. However, these effects were very subtle due to the distance to the water surface. 

The specific instructions for interaction were included because in Study 1 we received feedback indicating that more specific instructions would have helped in observing the motions of the pull tabs. In addition, since the subjects were interacting in a large-scale urban environment, there were countless opportunities for "plausibility noise" which we wanted to avoid (such as subjects expecting logs to realistically break windows, dent cars, knock over tables, and so on). By giving specific instructions, we aimed to ensure that the subjects' responses were based on the motions of the logs only.


\begin{figure}[h!]
\begin{center}
\includegraphics[width=10.0cm]{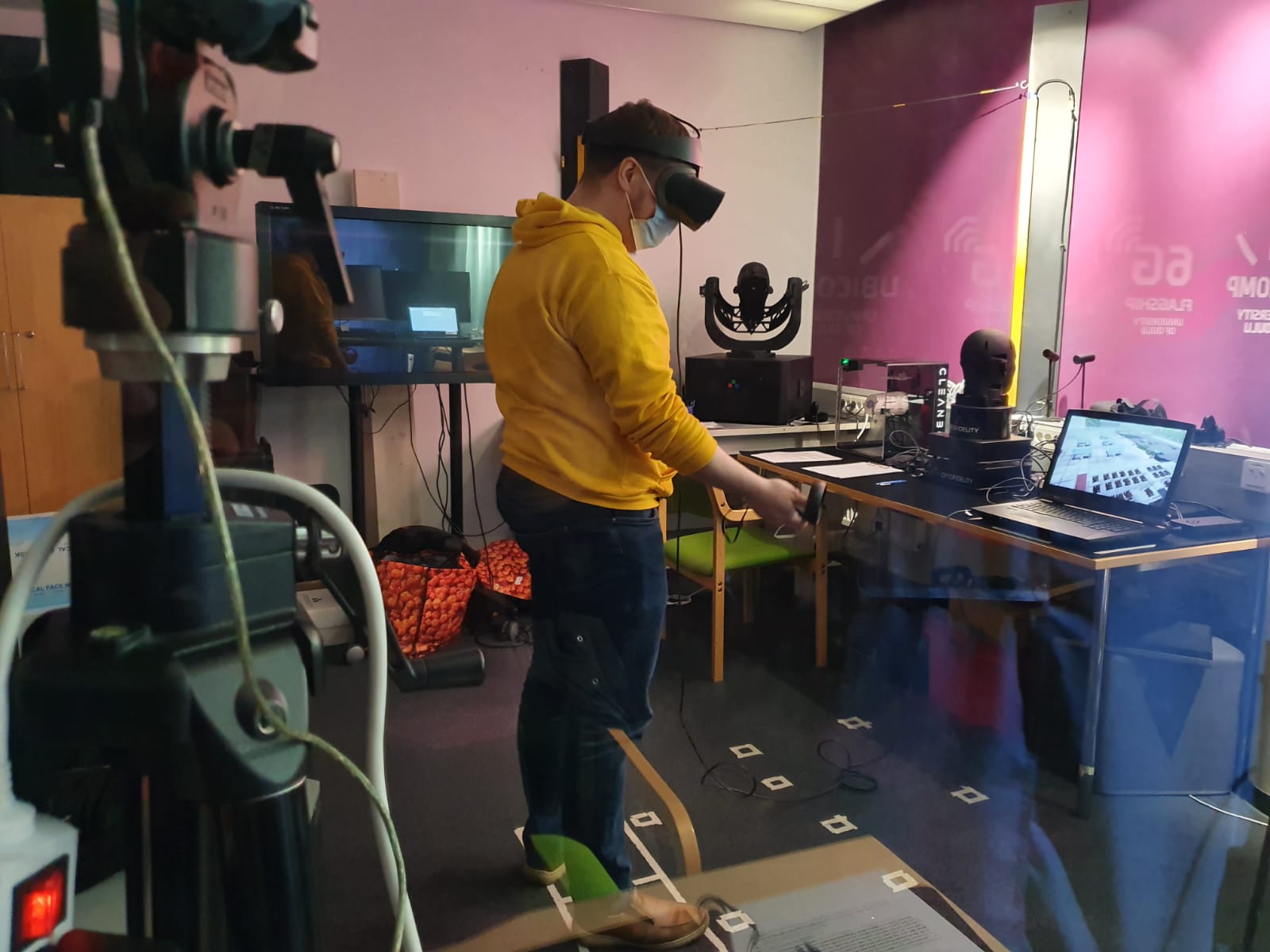}
\end{center}
\caption{A photo of the research space during Study 2 taken behind a see-through barrier. The same research space was used in both studies.}\label{lab}
\end{figure}

\subsubsection{Participants}
Similarly to Study 1, the experiment was carried out as a gender-counterbalanced within-subject experiment, this time with 40 participants (20 males and 20 females). Three participants were excluded due to failure to follow instructions or not giving their consent for data use. We did not allow people who had already participated in Study 1 to participate again to keep subjects naive for the purpose of the experiment. The subjects' ages ranged from 20 to 57, with mean and median ages being 26 and 25, respectively. The standard deviation of the ages was 6.0. Each participant was remunerated with a movie ticket worth 10 euros.

\subsubsection{Experimental Procedure}
Apart from COVID-19 related safety guidelines discussed below, the procedure was largely similar to Study 1. In Study 2, however, the subjects did not wear noise-cancelling headphones as there was no sound in the VE or from students in the laboratory hallways, and thus they were unnecessary. 

In Study 1, a researcher checked the HMD before each participant to ensure that the Oculus main menu or other anomalies were not present when starting the experimental apparatus. In Study 2, however, we asked the subject to report what he/she saw in the beginning of the experiment since we could not be in close proximity to the subjects to check ourselves. In addition to checking anomalies, this also allowed us to check whether the subject recognized the VE as the Oulu marketplace. After this, an instructions script was read for the subject. The script confirmed that the subjects were at Oulu marketplace, enlarged 10-fold to a size of a giant, and they were to throw and drop the logs placed on a container in front of them. At the end of the experiment, the subject was given her/his movie ticket.

The experiment in Study 2 was conducted during the COVID-19 pandemic, hence additional safety precautions were taken. At the time of the experiment, the regional state of the epidemic was at so-called "baseline level"\footnote{Defined by Finnish Institute of Healthcare as follows: "The baseline corresponds to the situation in Finland in the middle of the summer, 2020. The incidence of infections is low, and the proportion of endemic infections is small."}. Due to the relatively calm local state of the epidemic, it was possible to conduct temporary on-campus work as long as university safety guidelines were followed. 

The research space allowed for a maximum of two researchers who kept within safety distance to the participant. The researchers were also separated from the subject with a see-through barrier. The researchers wore safety masks, which were also offered for subjects.  The participants were instructed to operate the VR equipment by themselves during the experiment and a researcher intervened only if necessary (for example, in cases of Oculus room setup resetting). 

Virtual reality equipment was sanitized between each participant using a "Cleanbox"\footnote{https://www.cleanboxtech.com/} device. In addition, all equipment and surfaces were wiped with alcohol disinfectants. Researchers wore rubber gloves during the cleaning process and the experiments. The default face padding of Oculus Rift S was covered with a silicone hygiene cover for easier cleaning. In addition, the subjects were offered optional disposable paper face hygiene covers. The research space was air-conditioned and ventilated between subjects. Participants were also asked to use hand disinfectant available in the research space. Participants were asked to join the experiment only when feeling completely healthy. The research space can be seen in Fig. \ref{lab}.

\subsubsection{Questionnaires}
The questionnaires in Study 2 were kept mostly similar to Study 1, consisting of two forced choice questions (main questions 1 and 2), two open-ended questions (O1 and O2), a 7-point Likert questionnaire concerning log physics (L1-L5), and the extended SUS questionnaire \cite{slater_depth_1994,usoh2000using}. In addition, we added extra 7-point Likert-scale questions L6-L8 concerning the experience of being large and pseudohaptics.

Main questions 1 and 2 were identical to Study 1, except replacing "pull tabs" with "logs". Similar to Study 1, both main questions were followed by open-ended questions O1 and O2 stating \textit{"Why?"}.


Questions L1, L3, L4 and L5 were kept similar to Study 1 so that only "pull tabs" were changed into "logs." Since the wording of L2 in Study 1 was found to be problematic, we paraphrased it from \textit{"the speed of pull tabs when thrown (slow - fast)"} into \textit{"time of flight (slow - fast)"}.

The new questions L6-L8 assessed the feeling of size and the sensation of weight of the logs in both conditions. The questions were phrased as follows.

\begin{enumerate}
    \item [L6] \textit{During the experiences, did you feel more like a giant in a normal-sized city, or more like a normal-sized person in a miniature city? (normal-sized person, giant)}
    \item [L7] \textit{When picking up or holding the logs, did you feel a sensation of actual weight? (not at all, very much so)}
    \item [L8] \textit{The logs felt... (light, heavy)} 
\end{enumerate}

Similarly to Study 1, all questions were presented either in English, or Finnish, depending on the preference of the subject.

\subsection{Results}
\subsubsection{Hypotheses}
Again, majority of the subjects considered the \textit{movie physics} condition as the realistic one, but the expectations of the subjects were more mixed, however. For main question 1, 28 out of 40 (70\%) subjects chose \textit{movie physics}. As for response to the main question 2, 25 out of 40 subjects (63\%) considered that \textit{movie physics} matched their expectations better. Following the procedure in Study 1, we analyzed the frequencies of responses with a binomial test, and found their corresponding two-tailed p-values as $p=0.017$ and $p=0.154$. This indicates that the responses to main question 1 were significantly biased towards \textit{movie physics}, whereas responses to main question 2 were closer to a random distribution. These results confirm H1, but not H2. Most of the subjects clearly considered \textit{movie physics} as the more realistic condition, but their expectations were more evenly split between \textit{true physics} and \textit{movie physics}. Almost every subject mentioned recognizing the scene as the Oulu downtown marketplace. 


\subsubsection{Open-ended questions O1 and O2}
\begin{figure}[h!]
\begin{center}
\includegraphics[width=18cm]{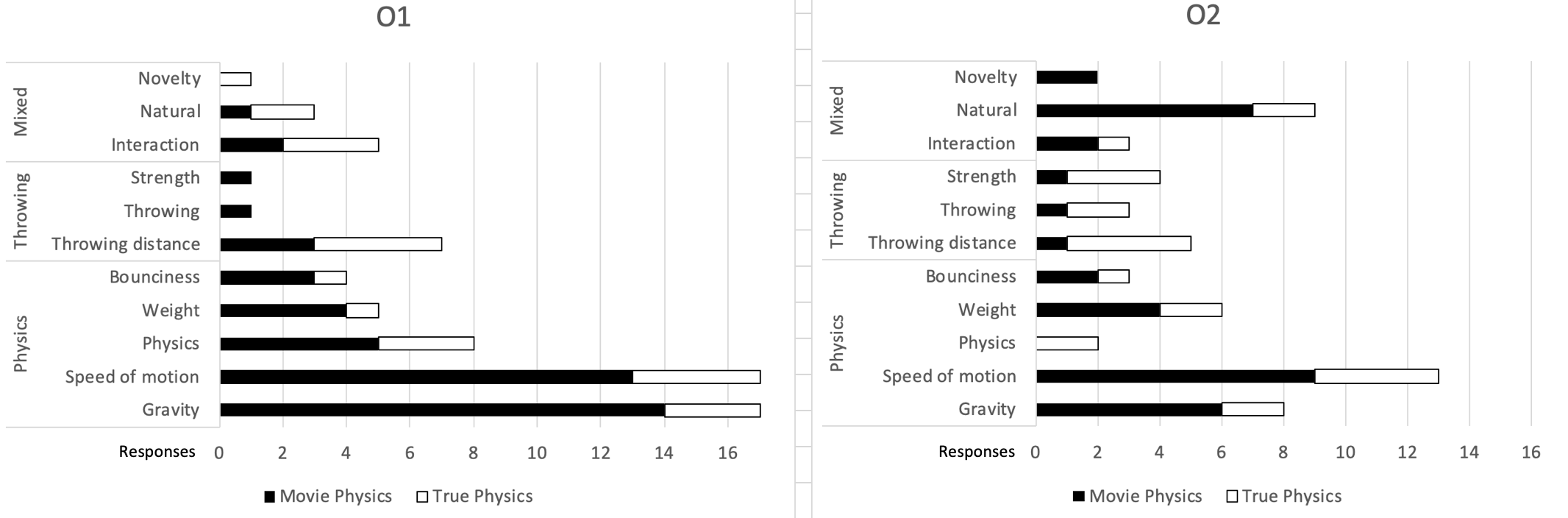}
\end{center}
\caption{Qualitative codes for responses O1 (\textbf{A}) and O2 (\textbf{B}) according to pereived realism in Study 2} \label{qualistack_separated}
\end{figure}

Thematic analysis using the inductive approach (\citealt{patton2005qualitative}) was used to analyze the open-ended questions. The responses were first coded by two independent researchers, after which the final codes were agreed upon. One subject did not respond to the open-ended questions. A summary of codes and their frequencies per each question can be seen in Fig. \ref{qualistack_separated}.



The responses indicate that majority of the subjects considered the motions of the logs as their primary reason of preference; the logs were either moving at a speed they did not feel was realistic, or were under the effect of abnormal gravity. This is especially true for the subjects that perceived \textit{movie physics} as more realistic. For the subjects choosing \textit{true physics}, the ability to throw logs especially far came up relatively more often than for \textit{movie physics} respondents. This could mean that these subjects considered a giant being capable of throwing the logs farther due to increased strength. However, similarly to Study 1, we did not simulate muscle strength per se; the ability to throw the logs far was due to the increased velocity the subjects were able to impart due to being scaled 10-fold. There was only one response to O1, in which \textit{strength} was specifically mentioned, whereas for O2, \textit{strength} came up in four responses. Examples of responses for O1 can be seen in Table \ref{O1examples}. For O2, examples can be seen in Table \ref{O2examples}. Distributions of qualitative codes in Study 2 can be seen in Fig. \ref{qualistack_separated}.


\begin{table}[]
\tiny
\centering
\caption{Examples of O1 responses (justification to main question 1) in Study 2}
\begin{tabular}{|p{3cm}|l|p{10cm}|}
\hline
\textbf {Codes} & \textbf {Preference} & \textbf{Response} \\
\hline
\hline
\textit{Gravity} & \textit{Movie physics} & \textit{"me being big should not affect the gravity of other objects"}\\
\hline
\textit{Speed of motion} & \textit{True physics} & \textit{"Not really sure, but when I picture a giant it feels like that way. Like things going on slow motion."}\\
\hline
\textit{Speed of motion, gravity} & \textit{Movie physics} & \textit{"It happened with normal speed/gravity"}\\
\hline
\textit{Physics, gravity} & \textit{Movie physics} & \textit{"Acceleration felt somewhat realistic, the latter felt like surface of the moon"}\\
\hline
\textit{Throwing distance} & \textit{True physics} & \textit{"when using a strong force, the logs was thrown far away, matching my expectation"}\\
\hline
\textit{Throwing distance} & \textit{True physics} & \textit{"If I were a giant, the logs would fly a little farther, which was highlighted in the second time"}\\
\hline
\textit{Novelty, physics, bounciness, interaction} & \textit{True physics} & \textit{"Everything felt new, not only that you were in VR in the first place, but also the point of view, which was of course higher than normal. It also felt like, in terms of physics, the logs were behaving more realistically in the first time, because I was handling them more carefully. On the second time I just dropped the logs from high up, and they were bouncing any which way".}\\

\hline
\textit{Gravity, weight, naturalness} & \textit{Movie physics} & \textit{"The gravity and motion of the logs felt more natural. In the other one, they were floating like feathers in space and were clearly lighter than real".}\\
\hline

\textit{Gravity, speed of motion, strength} & \textit{Movie physics} & \textit{"According to my own assumptions, objects would feel like they were moving more slowly in relation to myself if I were a giant, but the first time felt more like I was underwater. In my opinion, the second time was more real, even if it was a little fast-ish. I did feel as if I was stronger in the second time, though.".}\\
\hline

\end{tabular}
\label{O1examples}
\end{table}

\begin{table}[]
\tiny
\centering
\caption{Examples of O2 responses (justification to main question 2) in Study 2}
\begin{tabular}{|p{2.5cm}|p{2.5cm}|p{10cm}|}
\hline
\textbf {Codes} & \textbf {Preference} & \textbf{Response} \\
\hline
\hline
\textit{Naturalness, speed of motion} & \textit{True physics (different from O1)} & \textit{"This bias might be partly because of movies, but also in many real life videos, big objects fall "more slowly" when seen from afar. In the first version the logs were much more slower, which matched my expectations more."}\\
\hline
\textit{Naturalness, speed of motion} & \textit{Movie physics (different from O1)} & \textit{"Somehow faster motions felt more natural"}\\
\hline
\textit{Naturalness} & \textit{Movie physics} & \textit{"Logs felt more credible in the second experiment"}\\
\hline
\textit{Physics} & \textit{True physics} & \textit{"The second time matched my expectations more since the motions of the logs were more realistic"}\\
\hline
\textit{Speed of motion} & \textit{Movie physics} & \textit{"Because the logs acted as they should. A log in the real would not fall slowly."}\\
\hline
\textit{Interaction, bounciness, throwing} & \textit{Movie physics} & \textit{"I find it easy to grab and on throwing it was more realistic. When I drop the log it bounced back as well, making it more realistic. In second, I was also able to see the  log clearly when it was in air during the throw."}\\
\hline
\textit{Speed of motion} & \textit{Movie physics} & \textit{"Still the first one. I can not realistically think the world working in slow motion.".}\\

\hline
\textit{Speed of motion} & \textit{True physics} & \textit{"because of the speed when I drop the logs".}\\
\hline

\textit{Strength, throwing distance} & \textit{True physics} & \textit{"As a giant I would expect to be stronger, therefore being able to throw the logs further. ".}\\
\hline

\end{tabular}
\label{O2examples}
\end{table}


\begin{figure}[h!]
\begin{center}
\includegraphics[width=17.5cm]{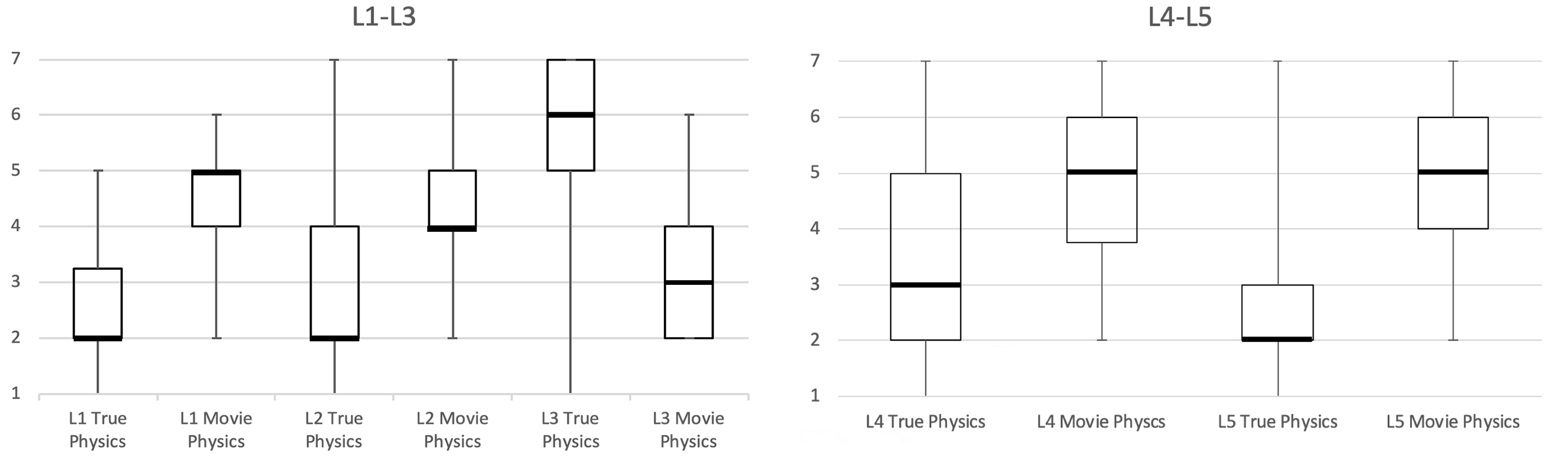}
\end{center}
\caption{Study 2 Likert responses L1-L3 (\textbf{A}) and L4-L5 (\textbf{B}) visualized as box plots, interquartile ranges and minimum and maximum responses. In (\textbf{A}), responses closer to 4 are perceived as closer to realism, whereas in (\textbf{B}) responses closer to 7 are perceived as closer to realism.}\label{fig:s2likert}
\end{figure}

\subsubsection{Likert Data}
Similarly to Study 1, we analyzed questions L1-L5 (\textit{falling speed, time of flight, distance when thrown, bounciness}  and \textit{gravity}) to get additional insight into the subjects' perceptions of the motions of the logs. In all questions, the respondents favored \textit{movie physics}, with the median and mode closer to 4 in L1-L3 and closer to 7 in L4 and L5. A Wilcoxon Signed Rank test showed that the responses were significantly different $(p < 0.05)$ between the conditions for all of the questions except L4 $(p = 0.4)$. A summary of the responses can be seen in Table \ref{fig:s2likert}. Box plots visualizing medians, interquartile ranges, as well as the minimum and maximum responses can be seen in Fig. \ref{fig:s2likert}.


\begin{table}[]
\small
\centering
\caption{Summary of Likert data in Study 2. Responses perceived closer to realism are emphasized in bold.}
\begin{tabular}{|l|l|c|c|c|}
\hline
\textbf {Question} & \textbf {Condition} & \textbf{Median} & \textbf{Mode} & \textbf{STD} \\
\hline
\hline
\textbf{L1:} & \textit{true physics} & 2 & 2 & 1.3 \\
Falling speed & \textit{movie physics} & \textbf{5} & \textbf{5} & 1.0         \\
\hline
\textbf{L2:} &  \textit{true physics} & 2 & 2 & 1.7 \\
Time of Flight & \textit{movie physics} & \textbf{4} & \textbf{4} & 1.1 \\
\hline
\textbf{L3:} &  \textit{true physics}  & 6 & 6 & 1.5 \\
Distance when thrown & \textit{movie physics}  & \textbf{3} & \textbf{4} & 1.0 \\
\hline
\textbf{L4:} & \textit{true physics}  & 3 & 2 & 1.7 \\
Bounciness & \textit{movie physics}  & \textbf{5} & \textbf{6} & 1.5 \\
\hline
\textbf{L5:} & \textit{true physics} & 2 & 2 & 1.7 \\
Gravity & \textit{movie physics} & \textbf{5} & \textbf{6} & 1.4 \\
\hline
\end{tabular}
\label{Likertdata}
\end{table}

\subsubsection{Presence Data}
Similarly to Study 1, we acquired self-reported presence data according to the SUS questionnaire. Thirty out of 40 subjects (75\%) had an SUS score higher than 0, indicating at least some level of presence. The median SUS score was 1. Again, we divided the subjects into groups of high presence (SUS score \textgreater~2) and low presence (SUS score \textless~3). The proportion of the high presence group was 16 out of 40 (40\%), whereas the low presence group consisted of 24 subjects (60\%). Seven subjects out of 16 (44\%) from the high presence group chose \textit{true physics}, whereas only five subjects out of 24 (21\%) from low presence group did the same. However, according to Fisher's exact test, this difference was not significant ($p > 0.05$), which means we can assume that belonging to either high or low presence group did not affect the response to main question 1.

\subsubsection{Own Size and Pseudohaptics Sensations}
We added three new questions L6-L8 to investigate the subjects' perception of his/her own size, as well as pseudohaptic sensations. 
It appears that although the subjects generally considered \textit{movie physics} as the more realistic condition, \textit{true physics} was able to more successfully convey the sensation of being large. The median and mode responses to L6, \textit{feeling of own size}, were 6 and 6, respectively, for the \textit{true physics} conditions. As for \textit{movie physics}, these responses were 4 and 2, respectively. This is somewhat supported by the responses to open-ended questions O1 and O2 (for example, subjects considering true physics more natural, see Table \ref{O2examples}). We found these differences to be statistically significant using the Wilcoxon Signed-Rank test ($p = 0.002$).

When inspecting the open-ended data from Study 1, we found a number of subjects mentioning the pull tabs feeling heavier in one condition or another. To investigate this further, we added new questions L7 and L8 to inquire about pseudohaptic sensations. However, these sensations were reported as very low in general. For L7, \textit{sense of actual weight}, the median and mode responses were 1.5 and 1 for \textit{true physics} and 2 and 1 for \textit{movie physics}, respectively. As for L8, \textit{logs felt light/heavy}, median and mode was 2 and 1 for \textit{true physics} and 3 and 1 for \textit{movie physics}. Out of 40 subjects, 4 and 6 subjects reported pseudohaptic sensations stronger than 4 out of 7 in the \textit{true physics} and \textit{movie physics} conditions, respectively. However, Study 2 might have been less appropriate to study the sensations of weight reported by subjects in Study 1, since in Study 2 the physics ranged from normal to perceptually slower, instead of vice versa. Again, using the Wilcoxon Signed-Rank test, we found statistical differences for L7 (\textit{physical sensation of weight}) to be insignificant ($p > 0.05$). However, there was a statistically significant difference ($p = 0.026$) in responses for L8 (the logs were light/heavy) . These results could be interpreted so that the subjects generally considered the logs simulated heavier in the \textit{true physics} condition, but failed to notice any differences regarding pseudohaptic sensations, however.

\subsubsection{Effect of Background, Presence and Perception of Own Size}
Similarly to Study 1, we analyzed the effects of subject background and self-reported presence on their preference on physics. This time around, we also added responses to L6 as variables \textit{own size true} and \textit{own size movie} to estimate whether subjects' perception of their own size (in essence, the extent of feeling like a giant) affected responses. Using the same categories and the same coding mechanisms as in Study 1 (this time with 22 subjects categorized having a background in  \textit{Natural Sciences and Engineering} and 17 subjects with a \textit{Social Science} background), we performed binary logistic regression analysis. The model explained 38\% of the variance (Nagelkerke's $R^2$) with 79.5\% overall accuracy. Although we found, similar to Study 1, that background, gaming or VR experience and self-reported presence did not affect responses, the variable \textit{own size movie} had a significant effect ($p = 0.041$). This finding indicates that \textit{true physics} respondents felt smaller specifically during the \textit{movie physics} condition. However, since the distributions of both \textit{true physics} and \textit{movie physics} responses were quite large, but the number of \textit{true physics} respondents was rather small, we would hesitate to put too much confidence in this implication until further evidence is found.

\subsection{Comparing Studies 1 and 2}
The percentage of subjects that chose \textit{movie physics} for main question 1 was 73\% in Study 1 and 70\% in Study 2. As for main question 2, these percentages were 91\% for Study 1 and 63\% for Study 2. We compared the results for main questions 1 and 2 from Studies 1 and 2 with Fisher's exact test. We found that responses to main question 1 were statistically similar ($p > 0.05$), whereas responses to main question 2 were different ($p = 0.003$). This suggests that a majority of similar proportions considered \textit{movie physics} more realistic in both studies. The proportions were largely different for main question 2. Although almost all subjects considered \textit{movie physics} as better matching their expectations in Study 1, only a statistically insignificant majority considered the same in Study 2.    

The results for Likert questions L1, L3, and L5 were very similar in Studies 1 and 2, consistently favoring \textit{movie physics}. Responses to L2 were very mixed in Study 1, which we attribute to bad wording of the question. In Study 2, the responses were more consistent and clearly favored \textit{movie physics}. In Study 2, the responses for L4 were mixed while in Study 1 \textit{movie physics} was preferred.

In both studies, we examined the effect of various contributing factors in an effort to gather additional insights for interpreting the results. In Study 1, we used background data as well as self-reported presence as predictors to main question 1. In Study 2, we also added two new variables \textit{own size true} and \textit{own size movie}. In Study 1, however, we did not find any significant predictors. In Study 2, a new variable, \textit{own size movie} came out as significant. 

If we compare the presence scores to those of Study 1, we can see that subjects in Study 2 experienced somewhat less presence. In Study 1, some presence (SUS score \textgreater~0) was experienced by 82\% of the participants with median SUS score being 3. In addition, the proportion of high and low presence groups were almost equal in Study 1 (53\% experiencing high sense of presence). In Study 2, 75\% responded with SUS score \textgreater~0 while the median SUS score was 1. The proportions of high and low presence groups were 40\% and 60\% respectively. However, despite these differences, the SUS scores for Study 1 (44 subjects) and Study 2 (40 subjects) were not statistically different (Mann-Whitney U test $p > 0.05$). Also, presence did not have a predictive capability on the preference of realism in either study.

\section{Discussion}
Our results demonstrate that we have identified a strong paradox concerning PSI in VEs in which the user has been scaled either up or down. However, this fits the definition of PSI: the plausibility illusion is more dependent on the expectations of the subjects than objective reality (\citealt{slater_depth_1994, skarbez2017survey}). We believe that this paradox has implications for VR and telepresence applications.

The proportion of the subjects that chose \textit{movie physics} in main question 1 was almost identical in Study 1 and Study 2. Close to a 3/4ths majority (73\% in Study 1 and 70\% in Study 2) chose \textit{movie physics} as the realistic representation. As for main question 2, the responses were quite different, however. In Study 1, 91\% of subjects considered \textit{movie physics} as matching their expectations more, whereas in Study 2 only 63\% of the subjects considered the same. It appears that realistically approximated physical phenomenon at a small scale was surprising for almost all subjects. However, many subjects considered \textit{true physics} to better match their expectations at a large scale, even if they actually regarded \textit{movie physics} as the realistic one. 

The purpose of open-ended questions regarding the reason why subjects rated one of the physics conditions being more realistic (O1) or matching their expectations better (O2), was first to confirm that the subjects gave their responses according to object motions and not other plausibility related factors, and second to give additional insights, for example regarding different responses to O1 and O2.

In Study 1, according to O1 and O2, almost all of the subjects considered their perception of realism to be related to the physics behavior of the tabs. In addition, a small number of subjects gave responses motivated by general interaction, including learning how to use the controllers correctly. A few secondary reasons or remarks were made referring to a scene object or other visual details. According to the responses to O2, most of the subjects preferring \textit{true physics} as the realistic one stated that during the experiment it was difficult to understand why the physics functioned the way that it did - the behavior of the tabs was still surprising even if they considered it realistic. 

As for Study 2, the reasons given by the subjects were also most often related to the behavior of the logs. Some exceptions included interaction (learning to use controllers or other interaction related issues) and novelty (for example, the experience being more overwhelming in the first part of the experiment). In Study 2, no visual aspects came up in the open-ended responses.

Whereas in Study 1, only one subject responded with \textit{movie physics} in O1 and \textit{true physics} in O2, this was the case for five subjects in Study 2. The most popular reason in these cases was the ability to throw the logs farther (3 responses). The other reason was that the slower motions somehow seemed more natural, even if unrealistic, as a giant (2 responses). Another difference to Study 1 was that for Study 2, subjects choosing \textit{true physics} in O1 usually gave the same response also to O2; the behavior of the logs at large scale was not surprising to the same extent as the behavior of the tabs in small scale. 


We used Likert scale rating questionnaires to gather additional insight into our findings. The questions focused on various dynamic properties of the objects so that we could more specifically pinpoint the effects of physics simulations on perceived realism. These responses indicated preferences towards \textit{movie physics} as well, with significant differences regarding the perceived realism of the object behavior (with the exceptions of question L2, \textit{speed when thrown,} in Study 1 and L4, \textit{bouncincess,} in Study 2). The Likert data further confirms that physically accurate representations of physics in abnormal scales are not inherently intuitive for VR users.


According to our results, accurate accelerations and falling speeds of objects were perceived as unrealistic. The distance that the subjects were able to throw the objects was seen mostly as too short in Study 1 and too long in Study 2. However, there were also responses in both studies that considered \textit{movie physics} to be too extreme. 

In Study 1, responses regarding the bounciness of the of the tabs indicated that subjects expected the tabs to behave as if they were enlarged 10-fold. In Study 2, however, the reactions to bounciness were much more mixed; even if median and mode responses preferred \textit{movie physics}, the responses were too mixed to cross the threshold of significance at $p = 0.05$. We believe the main reason for the difference for these responses is the scale. In Study 1, the tabs were not practically bouncing at all in the \textit{true physics} condition. In Study 2, however, the logs were bouncing in both conditions.

In Study 2, we inquired about the extent to which the subjects felt like a giant in a normal-sized city instead of a regular-sized person in a miniature city. We found a significant difference between the conditions, indicating the subjects in the \textit{true physics} condition felt larger. This may mean that even if the subjects did not generally believe the \textit{true physics} condition to be realistic, it succeeded better in providing the illusion of being large.

We inspected the effects of various aspects of the subjects' background on their responses to O1. It could be that the subjects with knowledge of physics, for example, might prefer the \textit{true physics} condition. However, we found no such effects in either of our subject groups. In addition, we did not find the self-reported level of presence (\citealt{slater_depth_1994}), either as SUS scores or by dividing subjects into groups of high and low presence, to affect the response to O1 in either study. In Study 2, we found a significant effect for the variable L6 \textit{feeling of own size - movie physics}. This suggests that the extent to which the subjects experienced the illusion of being a giant in the \textit{movie physics} condition had at least some effect on the subjects' perception of physics. However, the overall performance of the classifier was not very good, and the distributions of the subjects' responses were very large. For this reason, we believe further investigation is necessary before we can claim whether or not the extent of the small-scale or large-scale illusion affects the perception of physics.

In Study 2, we also studied pseudohaptic sensations experienced by the subjects and found that the overall extent of the sensations was very low. A handful of subjects reported strong pseudohaptic sensations. There was, as expected, a perceived difference regarding the overall weight of the logs between conditions. 

We found that the level of presence experienced by subjects in Study 2 was somewhat lower. However, as of now, we do not have evidence to claim whether the illusions of being small or large affected self-reported presence or whether, for example, the properties of the VEs used in the studies would explain these differences. 







\subsection{Implications}
\cite{slater_place_2009} discussed the role of conformity to expectations, prior beliefs and knowledge for causing and maintaining PSI. \cite{skarbez2020immersion} conceptualized the former as \textit{coherence}, the reasonable behavior of the VE, which, according to Skarbez, is related to PSI similarly as immersion is related to PI.

Looking at the results against this framework, we can see that in Study 1, \textit{movie physics} was clearly the reasonable behavior for subjects. Even if 27\% of subjects considered \textit{true physics} as real, only a handful of subjects considered it matching their expectations. Therefore according to the results of Study 1, realistic object behavior in small scale clearly violated coherence.

According to the results of Study 2, it is somewhat unclear which behavior is the coherent one, even if the results are somewhat pointing towards \textit{movie physics}. Although a significant majority did consider \textit{movie physics} as the realistic behavior, the expectations of subjects were matched almost even. Because of this mixed response to expectations, it is not straightforward to say, which model would yield good coherence in VEs.

If one was to design a multiscale VR application that would aim at maximizing coherence instead of realism, it would make sense simply to match the physics with the scale of the user, at least in small-scale applications. If the user is allowed to change scale, the physics behavior would follow similarly to Hollywood movies such as \textit{Honey I Shrunk the Kids} where object motions constantly change in speed from scene to scene according to perspective changes. In large scales, however, this type of behavior might lead to bad coherence. In addition, in mCVEs this approach would break since the physics model would not be able to feasibly accommodate all users' perspectives simultaneously during multi-user interaction.   

If realistic physics are intended, then users' expectations would have to be modified by some type of training so that realistic behavior does not come up as surprising. According to \cite{skarbez2020immersion} bad coherence in VEs, especially in relation to unexpectedly behaving environment, can lead to stress and discomfort. In addition, there might be cases where human interaction capabilities are reduced due to unexpected physics. Micro- and nano-evel robotics operations are an example of this \cite{sitti_microscale_2007}. For this reason we consider interaction at abnormal scales and perceptual training as important future research directions; even if users would expect realistic physics, their performance might still be affected.


Through recent advances in consumer VR hardware as well as sub-microscopic (\citealt{plisson_2d_2015}) and even atomic (\citealt{zheng_motioncor2:_2017}) level imaging techniques, it is possible that we will witness an increasing exploitation of scaled-down VR applications in the future. They could potentially include commercial systems such as teleoperated maintenance robots or commercial virtual design solutions at a microscopic scale. However, at this stage, it is unclear whether it would be intuitive for humans to operate at small scales, especially if it involves operating in the real world or with realistically simulated physics. As can be seen from our results, the perception of physical phenomena as a scaled-down entity is likely to be unintuitive for most. However, it was interesting to note that half of the subjects experienced a strong PI despite the apparent improbability of the experience of being doll-sized. As the scale of operation decreases, perceived frictions and accelerations increase, which has already been found problematic for humans in robotic micro- and nano-level operations (\citealt{sitti_microscale_2007}). As the scale decreases further, these perceived distortions amplify, and additional phenomena such as fluid dynamics and static electricity, come into play as well. Relative changes in the environment would also provide additional challenges in the physical domain. For example, a floor that is experienced as smooth at a regular scale might become bumpy and full of cracks. Grit and dirt might become actual obstacles for  navigation. Vibrations from passersby otherwise indistinguishable might feel like earthquakes. 
We also investigated the perception of physics at large scale. Study 2 enlarged the subjects 10-fold while giving them a similar interaction task. Although the users believed \textit{movie physics} as realistic similarly to Study 1, the expectations of users was much more mixed. We believe this finding might be useful for designing abnormal-scale VEs where PSI is more important than realism such as games. Realistic physics in small-scale interactions greatly violated the expectations of users while in large-scale slow, realistic motions sometimes seemed natural, even if ultimately unrealistic. 

We argue that our study opens up interesting avenues for future VR research. VR education has already been seen as a potential remedy for some issues of small-scale activities in the field of teleoperation (\citealt{millet_improving_2008}). Further research on the effect of perception-related mismatches on interaction and performance in various applications could yield interesting findings. Also, as of now, we do not know whether the body-scaling effect affects the perception of physics the same way it affects the perception of sizes and distances (eg. \cite{van_der_hoort_being_2011}). In both studies, we used virtual hands to provide a body-based size cue, but we did not investigate the effect the absence of these cues would have had. \cite{langbehn_scale_2016} found that groups of human avatars can override the dominant scale otherwise dictated by body-based size cues. Theoretically, this could have implications for perception of physics, as well.

\subsection{Challenges and Limitations}
Outliers in responses were L2 in Study 1 and L4 in Study 2.
Inspecting the distribution of responses in question L2 in Study 1, we see that the \textit{true physics} condition contains responses that are rather uniformly distributed in comparison to the \textit{movie physics} condition; the STD in the \textit{true physics} condition is twice as large as in the \textit{movie physics} condition. Whereas in responses to L2 the \textit{movie physics} condition was considered realistic (4, neither too fast nor two slow) by a vast majority, the \textit{real physics} condition received an almost equal number of responses between 2 (too slow) and 6 (too fast). We suspect that the uncharacteristic distribution of the responses might be due to a poor wording of L2 (\textit{The speed of pull tabs when thrown}). Although we tried to ask how the subjects perceived the time of flight of the tabs, it could be that subjects had other interpretations of the question resulting in inconsistent responses. Similar inconsistency was found in responses from both Finnish and English speaking subjects, so we do not think the confusion could be attributed to the specific wordings in either language. Rather, we speculate that some subjects thought we meant the speed of the tab in leaving their hand (resulting in short flight distance) upon throwing, and others thought we meant the speed that the tab moved through the air. Alternate interpretations could have resulted from misinterpreting the action of the tabs as having been caused by their own inability to throw the tabs correctly. We changed the wording of this question in Study 2 to \textit{"Time of Flight"}. 

In Study 2, L4 (the bounciness of the logs) received mixed responses. Although mean and mode responses preferred \textit{movie physics} similarly to other questions, the responses were overall more mixed. We consider bounciness as the most unrealistic aspect of Study 2, since we did not simulate splintering, or otherwise  breaking the log due to impact. We considered these aspects as confounding variables in a study that mainly focused on the perception of rigid body dynamics.

In Study 1, according to both verbal comments during the experiment as well as responses to questions O1 and O2, some of the subjects starting with the \textit{true physics} condition thought that the reason for their difficulty in throwing the tabs to a far distance was their own inability to use the controllers and not related to aspects of the VE. Although some subjects realized during the subsequent \textit{movie physics} condition that the behavior of the tabs was an experimental manipulation and not due to their own failure, there were still three subjects that stated as their main reason for preferring the \textit{movie physics} condition to be the fact that they had learned how to use the controllers. For subjects experiencing \textit{movie physics} first, there did not seem to be any ambiguity that the difference in the behavior of the tabs was related to the VE. Although a training session helping to learn the controllers might have been helpful, we believe that it could have introduced unwanted priming of the subjects regarding the expected behavior of physics. We received these types of responses far less in Study 2, which might be due to opposite behavior of objects when throwing.


Another obvious limitation is the fact that it is currently difficult to realistically simulate object mass in VR since subjects can feel only the weight of the controllers. Although we chose the soda can pull tabs for the task in Study 1 partly because of their light mass, there was some speculation among responses to O1-O2 on whether the weight of the object and/or simulated arm strength affected object manipulation. There were responses in Study 2 as well that considered throwing distance to be affected by the arm strength of the giant. However, simulating muscle strength in itself was not in the scope of either study. Human-scale arm motions were simply scaled either down or up by a factor of ten, which resulted in either very small or very large velocity imparted on the thrown object.

During a few experimental sessions, there were occurrences which could have broken presence or caused differences in the experiences of the participants. Two subjects in Study 1 became very active in the VE and accidentally bumped into furniture in the research space. In Study 2, one subject accidentally stepped on the HMD cord during the session. In Study 1, a physics engine bug caused a single tab to land in an unrealistic orientation during the \textit{true physics} condition for two subjects. For one subject trying to throw the tab with two hands, a bug caused the tab to catapult unrealistically far. We are not sure to what extent the subjects noticed these bugs or if it affected their responses. In Study 2, we did not observe physics bugs as obvious as those seen in Study 1. This might be partly due to the instructions for object manipulation being stricter and the scale of the objects being less prone for errors in the physics engine. Even still, we cannot guarantee that the bounciness of the logs was realistic at all times.


Additionally, although we tried to keep the visual appearances of the two conditions as similar as possible in Study 1, the differences in the VE scale in the UE to simulate the two types of physics led to very subtle differences in their brightness. This deficiency was fixed in Study 2 by simulating human-scale and giant-scale physics by manipulating gravity instead of scaling the scene objects.


Finally, there were subjects who were not always paying close attention to the flying or falling characteristics of the tabs, or did not wait until the instructions were read in their entirety. This limitation was somewhat alleviated in Study 2 due to the stricter instructions given to the subjects.


\section{Conclusion and Future Work}
In this paper, we studied a phenomenon regarding the plausibility of physical interactions for scaled-down and scaled-up users in normal-sized VEs; when users interact with physically simulated objects in a VE where the user is scaled 10-fold smaller or larger from a regular human scale, there is a mismatch between expected physics and the accurate approximation of physics at that scale. A similarly sized and significant majority of both scaled-down and scaled-up subjects judged rigid-body dynamics close to human-scale realistic instead of what would be the correct approximation of realism at the resized scale the subjects were on. Almost all subjects at a small-scale considered rigid-body dynamics at that scale to be surprising, while the expectations of large-scale subjects were more mixed. We argue that these findings open many interesting avenues for future research regarding mCVEs, scaled-down user VR applications in general, as well as telepresence and teleoperation taking place at a modified scale. In addition, our findings can prove useful to designers of VR applications utilizing abnormal scales, who wish to maintain PSI, or who are seeking to find a trade-off between PSI and realism.



In the future, we intend to study the body scaling effect and its influence on interactions with physically simulated objects. In addition, we will investigate interaction, performance, and perceptual training at abnormal scales. We will consider scales smaller than 1 order of magnitude since we expect them to provide even greater plausibility mismatches in physical interactions. We will also seek to confirm the existence of our finding outside VR, for example using robotic teleoperation or telepresence at small scale. Moreover, we will seek further evidence as to whether the extent of the illusion of being small or being large affects the perception of physics. 






\section*{Conflict of Interest Statement}

The authors declare that the research was conducted in the absence of any commercial or financial relationships that could be construed as a potential conflict of interest.



\section*{Funding}
This work was supported by the PIXIE project (331822) as well as the PERCEPT project (322637) funded by the Academy of Finland. This work was also supported by the COMBAT project (293389) funded by the Strategic Research Council at the Academy of Finland  as well as the HUMORcc project (6926/31/2018) funded by Business Finland.

\section*{Acknowledgments}
This paper extends the previous work by \cite{pouke2020plausibility} titled \textit{"The Plausibility Paradox For Scaled-Down Users in Virtual Environments"} originally published in the IEEE Conference on Virtual Reality and 3D User Interfaces (IEEE VR 2020). Parts of the original article are reused with permission. The authors also wish to thank all the subjects for their participation in this study.



\bibliographystyle{frontiersinSCNS_ENG_HUMS} 
\bibliography{ppresized}

\end{document}